\documentclass[preprint,12pt,number]{elsarticle}

\usepackage{amssymb}
\usepackage{amsmath}
\usepackage{mathtools}
\usepackage{pdflscape}
\usepackage{color}
\usepackage{tikz}
\usepackage{pstool}
\usetikzlibrary{positioning,arrows.meta,calc}
\usepackage[breaklinks=true]{hyperref}
\biboptions{sort&compress}

\usepackage[a4paper,margin=2.5cm]{geometry}

\newcommand{\B}[1]{\boldsymbol{#1}}

\newcommand{\mr}[1]{\mathrm{#1}}
\newcommand{\dA}{\mr{d}A}
\newcommand{\dV}{\mr{d}V}

\newcommand{\norm}[1]{\left| \left| #1 \right| \right|}

\renewcommand{\div}[1]{\mr{div}\left( #1 \right)}

\renewcommand{\exp}[1]{\mr{exp}\left( #1 \right)}

\journal{arXiv}

\begin{document}
\begin{frontmatter}

\title{Finite Element Modeling of Fracture in Brittle Foods with Physically Consistent Parameter Calibration: A Practical Case Study on Cooled Dark Chocolate}

\author{Patrick Kurzeja\corref{c1}}
\ead{patrick.kurzeja@tu-dortmund.de}
\author{Felix Rörentrop}
\author{Kai Langenfeld}
\author{Jörn Mosler}

\affiliation{organization={Institute of Mechanics, TU Dortmund University},
             addressline={Leonhard-Euler-Str.~5},
             city={Dortmund},
             postcode={44227},
             country={Germany}}

\cortext[c1]{Corresponding author}

%% Abstract
\begin{abstract}
Fracture in food can be undesired during production and transportation but also intended during cutting, grinding or mastication. The associated peak load, the deformation at failure and the resulting fragment geometry are all product properties. This work condenses the modeling of such fracture into a physically rigorous recipe that considers the practical constraints imposed by finite element simulations where necessary and applies idealized theoretical assessments where applicable. Cooled dark chocolate serves as a reproducible model material. Two well-established modeling frameworks for Griffith-type brittle fracture are assessed at an affordable spatial discretization: a sharp-interface and a diffuse phase-field approach. The assessment covers the choice of numerical framework, its discretization and a staged identification of the model parameters. Each parameter is identified from the loading stage that governs it: elastic parameters from pre-crack tension and compression, critical energy release rate from three-point-bending peak force, and the phase-field width last. The phase-field width interferes with the structural response, crack curvature and stress fields, depending on its lower bound set by the affordable spatial resolution. Elastic slope, peak force and crack path therefore cannot be met practically by a single parameter set. The objective must be fixed for the calibration of the model parameters and reported together with the numerical parameters and the boundary value problem.
\end{abstract}

%% Keywords
\end{frontmatter}

\clearpage
%%%%%%%%%%%%%%%%%%%%%%%%%%%%%%%%%%%%%%%%%%%%%%%%%%
%%% Introduction %%%
%%%%%%%%%%%%%%%%%%%%%%%%%%%%%%%%%%%%%%%%%%%%%%%%%%
\section{Introduction}
\label{sec:intro}

\subsection{Mechanical damage and fracture in foods: motivation and classification}

Mechanical damage in foods can occur during all steps of harvesting, processing, packaging and transport \citep{li_quantitative_2014,al-dairi_mechanical_2022,hill_pests_2003,gutierrez_stateart_2017}, impairing food quality and shelf-life. Damage may however also be inevitable, for example, in cutting processes \citep{boisly_experimental_2016,ageev_experimental_2021} or consumption \citep{swackhamer_fracture_2019,beckett_science_2019}. A precise and reliable prediction of the damage behavior is hence an important part of the design of both the products and the processing route \citep{rezende_mixture_2015,biswas_physical_2017}. From the perspective of food products, different damage descriptions are found, for instance, by location (e.g., abrasion or puncture \citep{li_quantitative_2014}), by load type (e.g., vibrations \citep{al-dairi_mechanical_2022} or freezing \citep{vanderSman_phasefield_2026}) or by length scales (e.g., cell damage in carrots \citep{ando_mechanical_2023} or tomatoes' locular gel tissue \citep{li_internal_2013}). Damage of food and packaging can cause loss of water, texture, nutrition, flavor, stability and infection resistance, among others, and undoubtedly results in substantial waste \citep{Liu_mechanical_2025,uhlig_food_2025,Wohner_Packaging_2019}. This clearly motivates the need for accurate and reliable predictions as one important tool to identify damage-prone loading conditions and to eventually reduce wastage along the food supply chain.

Such predictions rely on mathematical and numerical models for the mechanical damage process. These models allow for simulations that efficiently complement experimental and analytical approaches along the entire processing route, for instance in food engineering, cutting \citep{boisly_experimental_2016,ageev_experimental_2021}, mastication \citep{yven_role_2010}, frozen foods \citep{vanderSman_phasefield_2026}, micro-aeration in chocolate \citep{bikos_effect_2021,samaras_multiscale_2024}, fracture in toffee \citep{dammass_ratedependent_2024} and oral processing and teeth cleaning devices \citep{skamniotis_computer_2019}. 
For mathematical descriptions and simulations of food products on a length scale of mm to cm, it is important to distinguish two types of damage: ductile and brittle. Ductile damage is typically associated with plastic deformations. This can be relevant, for instance, during tomato compression \citep{li_internal_2013} or groove forming in food can lids \citep{boers_operator-split_2005}. Brittle damage, in contrast, is associated with elastic deformations. It can often occur instantaneously, without prior warning or noticeable (macroscopic) deformation. Examples include cracking in biscuits \citep{saleem_material_2005} or nuts \citep{swackhamer_fracture_2019}. Low temperatures, low humidity and high numbers of load cycles at small amplitudes moreover support brittle behavior.
The focus of this work is determined by brittle fractures. Choosing brittle behavior allows a better focus on the damage-related peculiarities, while many observations of this study also transfer to plasticity-driven ductile damage.

\subsection{Scope: Practical and physically rigorous modeling recipe for fracture in brittle foods}

The main scope of this work is to condense the modeling of brittle fracture in
food into a practical recipe, covering the choice of the numerical framework, its
discretization, the treatment of crack nucleation and propagation, and a physically consistent identification of the associated model parameters. Two well-established finite element frameworks are assessed for
this purpose, a sharp-interface and a diffuse phase-field description. Cooled
dark chocolate serves as a reproducible model material. The emphasis lies
deliberately on what is achievable in practice rather than on asymptotic
guarantees. Theoretical convergence results for idealized sharp-interface limits are used for the present assessment \citep{ambrosio_approximation_1990,bourdin_numerical_2000,bourdin_numerical_2007,bourdin_variational_2008}, but often neither reached nor affordable in an applied simulation \citep{linse_convergence_2017,may_numerical_2015,zhang_numerical_2017,nguyen_choice_2016}. Therefore, the spatial resolution of the present finite element simulations is set by a computationally affordable element size $h$, i.e., the edge length of the finite elements into which the specimen is subdivided. 

The main highlights of this investigation are:
\begin{itemize}
\item a modeling recipe for brittle fracture in food that assesses both a sharp-interface and a phase-field framework;
\item a staged identification of the elastic parameters, the critical energy release rate and the phase-field width;
\item an assessment of their calibration being tied to an affordable finite-element mesh size;
\item new experimental data for cooled dark chocolate, comprising tension, compression and bending tests.
\end{itemize}

Three quantities serve as modeling targets and as measures of agreement between simulation and experiment. They are traceable from comparatively simple experiments, and each of them matters in food engineering practice:
\begin{itemize}
\item the \emph{elastic slope} of the force-displacement response, which governs the deformation under sub-critical loads and is linked to sensory attributes such as hardness \citep{bikos_destructive_2023};
\item the \emph{peak force}, which defines the threshold for unintended fracture during handling, packaging and transport \citep{li_quantitative_2014,al-dairi_mechanical_2022} and the force required for intended fracture, e.g., during cutting \citep{boisly_experimental_2016} or mastication \citep{swackhamer_fracture_2019,skamniotis_computer_2019}; together with the elastic slope it determines the \emph{peak displacement}, which is decisive whenever a process is displacement-controlled, as in grinding and kneading;
\item the \emph{crack path}, which determines the fragment geometry and thereby the further processability, the appearance and the perceived quality of the product.
\end{itemize}

Measuring all three targets on one material calls for specimens that are
reproducible, geometrically flexible and cheap enough to be discarded in
numbers. Cooled dark chocolate meets these requirements. It behaves
homogeneously and isotropically on the mm-to-cm scale considered here, it can be
cast into non-standard geometries with predefined crack paths at low
manufacturing effort, and melting and re-casting make the
material reusable. This also makes it a convenient surrogate and the material chosen for the present case study. Alternatives are more restrictive; for example, nuts limit the available geometries and frozen foods or icing add sub-zero handling,
splintering or dusting. The contribution of this work is therefore twofold. It
applies both frameworks to a food material, and it assesses their fundamental
model parameters against a new experimental data set comprising tension,
compression, three-point-bending and plate-bending tests.

Elastic slope, peak force and crack path are likewise established reference quantities for non-food materials used in storage, tableware and packaging, e.g., for concrete \citep{Winkler2004}, technical ceramics \citep{Danzer_Fracture_2008} and structural glass \citep{wang_post-fracture_2020}. These targets require different model ingredients and the proposed recipe for the present case study is outlined in Fig.~\ref{Fig:Recipe}. The paper follows this sequence. Section~\ref{sec:frameworks} introduces the two frameworks and their parameters (basis for Steps~1 and 2). Section~\ref{sec:experimentalsetup} presents the new experimental data for cooled dark chocolate and the elastic parameters (Step~3). Section~\ref{sec:3pbending} identifies $g_c$ and $l_\varphi$ from three-point bending (Steps~4--5, based on specifications of Steps~1--2), and Sec.~\ref{sec:platebending} confronts them with the free crack path of a plate (Step~6). Section~\ref{sec:recipe} condenses the findings into the recipe, and Sec.~\ref{sec:conclusion} concludes.

\tikzset{
  rcpbox/.style={draw, rounded corners=2pt, line width=0.4pt, inner sep=6pt,
                 text width=11.4cm, align=left, font=\small}
}

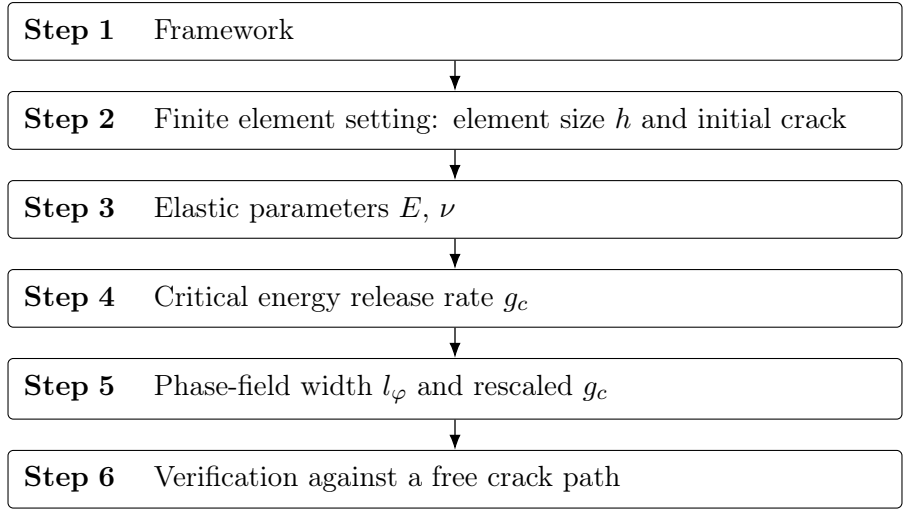
\begin{figure}[h]
\centering
\begin{tikzpicture}[node distance=4mm,
                    arr/.style={-{Latex[length=2mm]}, line width=0.5pt}]

\node[rcpbox]              (s1) {\textbf{Step 1} \quad Framework};
\node[rcpbox, below=of s1] (s2) {\textbf{Step 2} \quad Finite element setting: element size $h$ and initial crack};
\node[rcpbox, below=of s2] (s3) {\textbf{Step 3} \quad Elastic parameters $E$, $\nu$};
\node[rcpbox, below=of s3] (s4) {\textbf{Step 4} \quad Critical energy release rate $g_c$};
\node[rcpbox, below=of s4] (s5) {\textbf{Step 5} \quad Phase-field width $l_\varphi$ and rescaled $g_c$};
\node[rcpbox, below=of s5] (s6) {\textbf{Step 6} \quad Verification against a free crack path};

\foreach \a/\b in {s1/s2, s2/s3, s3/s4, s4/s5, s5/s6}
  \draw[arr] (\a) -- (\b);

\end{tikzpicture}
\caption{Modeling recipe for brittle fracture in foods used for cooled dark chocolate in this case study. This process can naturally be an iterative one as numerical and experimental settings are adjusted during the workflow, for example, to cover a wider parameter space or to increase sensitivity.}
\label{Fig:Recipe}
\end{figure}

%%%%%%%%%%%%%%%%%%%%%%%%%%%%%%%%%%%%%%%%%%%%%%%%%%
%%% 2 Numerical frameworks
%%%%%%%%%%%%%%%%%%%%%%%%%%%%%%%%%%%%%%%%%%%%%%%%%%
\section{Numerical frameworks and model parameters}
\label{sec:frameworks}
 
\subsection{Griffith-type fracture and sharp vs.~diffuse interface approaches}
Crack growth in brittle fracture is commonly captured by a Griffith-type criterion based on energy release rates \citep{griffith_vi_1921,bourdin_variational_2008}. A crack propagates once sufficient elastic energy is available to create new fracture surfaces in a food product, e.g., when bitten or broken during processing. Available models and technical implementations differ widely in computational cost and flexibility. Sharp-interface descriptions represent the crack as a discrete surface embedded in the body, whereas diffuse descriptions replace that surface by a damage field over a band of finite width. Both are represented here: a classical sharp-interface model and, as the diffuse counterpart, a phase-field model \citep{Langenfeld2022,langenfeld_curvature_2023}. This choice is not a judgment of the alternatives, each of which has its own range of applications. The extended finite element method (XFEM), for instance, retains a sharp crack but embeds it into the elements via enrichment functions, which allows for freely evolving crack paths at the price of additional numerical complexity \citep{belytschko_elastic_1999,moes_finite_1999}. Eigenerosion approaches introduce their own numerical length scale and control parameters \citep{schmidt_eigenfracture_2009,pandolfi_eigenerosion_2012}. Moreover, adaptive approaches modify the numerical mesh or the model parameters during crack propagation \cite{phansalkar_spatially_2022} and cohesive zone models add a strength to the fracture energy via traction-separation laws and have been applied to chocolate \citep{gomez_fracture_2022}.
 
\begin{figure}[h]
  \centering
    \includegraphics[width=0.9\textwidth]{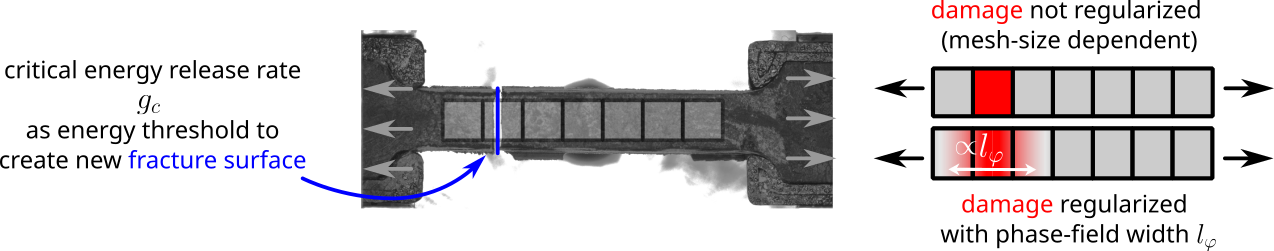}
    \caption{Illustration of the two key model parameters against a uniaxial tension test of cooled dark chocolate: the critical energy release rate $g_c$ (left) and the phase-field width $l_\varphi$ (right).}
    \label{fig:sketch_model_parameters}
\end{figure}

This section provides the basis for Steps~1 and~2 of the recipe: the two frameworks and their most important parameters. The elastic parameters are Young's modulus $E$ and Poisson's ratio $\nu$. The two fracture-related parameters in the focus of this work are the critical energy release rate $g_c$ and the phase-field width $l_\varphi$ (Fig.~\ref{fig:sketch_model_parameters}). Inertia and viscous effects are neglected, so that time merely parametrizes the loading history.
 
\subsection{Sharp-interface approach and the critical energy release rate}
\label{ssec:sharpinterface}
When a brittle food such as a crisp cracker is bent, elastic energy accumulates until advancing a crack releases more stored energy than is needed to create the new fracture surfaces, at which point the cracker snaps. The variational approach of \citet{Francfort1998} formalizes this competition. For a body $\mathcal B \subset \mathbb R^d$ (in spatial dimension $d = 2$ or $3$) that contains a crack set $\Gamma$, the total energy reads
\begin{flalign}
\mathcal E(\B u, \Gamma)=\int_{\mathcal B\setminus \Gamma} \psi_0(\B\varepsilon(\B u))\dV + g_c\,\mathcal H^{d-1}(\Gamma) - \int_{\mathcal B} \rho\, \B b \cdot \B u \dV - \int_{\partial \mathcal B_t} \B t\cdot \B u \dA .
\label{eq:total_energy}
\end{flalign}
The four terms represent the elastic strain energy of the intact material, the fracture energy, and the potentials of the body forces and surface tractions. Here, $\psi_0(\B\varepsilon)=\frac{1}{2}\,\B\varepsilon:\mathbb{E}:\B\varepsilon$ is the elastic strain-energy density with the linearized strain $\B\varepsilon(\B u) = \operatorname{sym}(\nabla \B u)$ and the elasticity tensor $\mathbb E$, $\rho$ is the mass density, $\B b$ the body force per unit mass and $\B t$ the traction prescribed on $\partial\mathcal B_t$. The Hausdorff measure $\mathcal H^{d-1}(\Gamma)$ equals the crack length in 2D and the crack surface area in 3D.
 
The material parameter
\begin{equation}
    g_c: \quad \text{the critical energy release rate}
\end{equation}
is the energy required to create a crack of unit surface area (unit: J\,m$^{-2}$) \citep{griffith_vi_1921}. It quantifies the resistance against crack growth and directly governs structural responses such as peak forces. For linear elastic materials, it is related to the fracture toughness via Irwin's relation $K_{\mathrm{I}c} = \sqrt{E'\,g_c}$ with $E' = E$ for plane stress and $E' = E/(1-\nu^2)$ for plane strain. Measured values may differ between opening (Mode~I) and shear (Modes~II and III) loading, which is commonly attributed to crack kinking, crack-interface friction, surface roughness and process-zone effects \citep{Hutchinson1992,Suresh1990,Erdogan1963}. Both frameworks used here employ a single, mode-independent $g_c$, which is adequate for the present tension-dominated cracking. In contrast to $g_c$, the energy release rate $G$, i.e., the energy released per unit crack extension, and the stress intensity factors $K_i$ depend on geometry and loading and can change as the crack grows.
 
For the numerical sharp-interface approach, the load is applied in increments, and the displacement field $\B u_n$ and crack set $\Gamma_n$ at time $t_n$ follow from the incremental energy minimization 
\begin{flalign}
(\B u_n, \Gamma_n)=\arg \min_{\B u \in \mathcal U_n(\Gamma),\, \Gamma} \mathcal E_n(\B u, \Gamma) \quad \text{s.t.} \quad \Gamma_{n-1} \subseteq \Gamma_n.
\label{eq:incremental_min}
\end{flalign}
$\mathcal E_n$ is the energy~\eqref{eq:total_energy} at time $t_n$. $\mathcal U_n(\Gamma)$ contains the displacement fields that satisfy the prescribed boundary conditions on $\partial\mathcal B_u$ and may jump across $\Gamma$. The crack advances from former $\Gamma_{n-1}$ to its new domain $\Gamma_{n}$ at time $t_n$. The constraint $\Gamma_{n-1} \subseteq \Gamma_n$ enforces irreversibility, i.e., the crack is not allowed to heal in the present analysis. For a crack advancing continuously along a given path, the stationarity of \eqref{eq:incremental_min} with respect to the crack area eventually recovers Griffith's criterion: the crack propagates once the energy release rate reaches $g_c$. The present numerical implementation imposes the crack along finite element edges in the simulation by doubling the nodes of the cracked element edge. Furthermore, the algorithm employs a so-called local energy minimization. This implies that a crack propagates if the driving force equals the critical energy release rate. By way of contrast, global energy minimization can lead to cracks if the driving force is significantly lower. Further details on the numerical implementation are provided in \ref{app:numerical_implementation} and \cite{knees_computational_2013}.

\subsection{Phase-field approach and phase-field width}
\label{ssec:phasefield}
The phase-field approach replaces the crack set $\Gamma$ by the phase-field variable $\varphi$, with $\varphi = 0$ in intact and $\varphi = 1$ in fully broken material \citep{Francfort1998,bourdin_numerical_2000,Miehe2010}. In the transition zone $0 < \varphi < 1$, the crack is smeared over a band of finite width with gradually degraded stiffness. Since $\varphi$ is an additional field in the entire domain and not fixed to element edges, cracks can ideally propagate and branch independently of the mesh topology. This makes the phase-field approach particularly suitable for free-discontinuity problems \cite{ambati_review_2015}, in which the crack path is not known in advance.

Following the sharp-interface energy~\eqref{eq:total_energy} term by term, the total energy becomes
\begin{flalign}
\mathcal E_{l_\varphi}(\B u, \varphi)=\int_{\mathcal B} f(\varphi)\,\psi_0(\B\varepsilon(\B u))\dV + g_c \int_{\mathcal B} \gamma(\varphi, \nabla\varphi) \dV - \int_{\mathcal B} \rho\, \B b \cdot \B u \dV - \int_{\partial \mathcal B_t} \B t\cdot \B u \dA .
\label{eq:total_energy_pf}
\end{flalign}
The load potentials are identical to Eq.~\eqref{eq:total_energy}, and the first two terms are the diffuse counterparts of the elastic and the fracture energy. In the elastic energy, the integration over the intact body $\mathcal B \setminus \Gamma$ is replaced by an integration over the entire body, weighted by the degradation function
\begin{equation}
f(\varphi) = c_0 + \left[ 1 - c_0 \right] \left[ 1 - \varphi \right]^2 .
\label{eq:Phasefield_DegradationFunction}
\end{equation}
It decreases from one in intact material (which is almost everywhere) to the small residual stiffness $0 < c_0 \ll 1$ in broken material, which keeps the finite-element equations regular. The strain-energy density $\psi_0$ is the same as in Sec.~\ref{ssec:sharpinterface}, and the stress follows as $\B\sigma = f(\varphi)\,\mathbb E : \B\varepsilon$. In the fracture energy, the crack surface measure $\mathcal H^{d-1}(\Gamma)$ is replaced by the volume integral of the crack surface density
\begin{align}
\gamma(\varphi, \nabla\varphi) = \dfrac{1}{2\,l_\varphi}\left[\varphi^2 + l_\varphi^2\,\nabla\varphi\cdot\nabla\varphi\right] ,
\label{eq:Phasefield_CrackDensity}
\end{align}
while $g_c$ retains its meaning as the energy per unit crack surface. This introduces the second key parameter,
\begin{equation}
    l_\varphi: \quad \text{the phase-field width},
\end{equation}
of unit length. It weights the gradient term in $\gamma$ and thus sets the width of the transition from $\varphi = 0$ to $\varphi = 1$ (Fig.~\ref{fig:sketch_model_parameters}, right). For a fully broken long bar, the optimal profile $\varphi(x) = \exp{-|x|/l_\varphi}$ decays to about 5\,\% within $3\,l_\varphi$ and yields $\int_{-\infty}^{\infty}\gamma\,\mr{d}x = 1$ independently of $l_\varphi$, i.e., the band carries the fracture energy of exactly one unit crack surface. For $l_\varphi \to 0$, the energy~\eqref{eq:total_energy_pf} $\Gamma$-converges to the Griffith-type energy~\eqref{eq:total_energy} under suitable assumptions \citep{ambrosio_approximation_1990,bourdin_numerical_2000}. The gradient term moreover regularizes the mathematical problem to become well-posed and prevents damage from localizing in single elements \citep{Langenfeld2022}. The phase-field width $l_\varphi$ is therefore initially introduced not as a material parameter but as a numerical parameter that ideally vanishes for practically sharp cracks.
 
In analogy to Eq.~\eqref{eq:incremental_min}, the displacement and phase fields at time $t_n$ follow from the incremental energy minimization
\begin{flalign}
(\B u_n, \varphi_n)=\arg \min_{\B u \in \mathcal U_n,\, \varphi} \mathcal E_{l_\varphi,n}(\B u, \varphi) \quad \text{s.t.} \quad \varphi_{n-1} \leq \varphi_n .
\label{eq:incremental_min_pf}
\end{flalign}
Details on the present numerical implementation are provided in \ref{app:numerical_implementation} and \citep{knees_computational_2013,roerentrop_etal_2026}. In contrast to the sharp-interface case, the set $\mathcal U_n$ contains continuous displacement fields satisfying the boundary conditions on $\partial\mathcal B_u$, since the displacement jump is replaced by the degraded band. The pointwise constraint $\varphi_{n-1} \leq \varphi_n$ is the counterpart of $\Gamma_{n-1} \subseteq \Gamma_n$ and enforces irreversibility, while the upper bound $\varphi \leq 1$ is satisfied by the minimizers automatically. For comparison, formulations based on energy rates \citep{Petryk1982, Comi1995, Ortiz1999, Carstensen2002, Miehe2010} lead to the same incremental problem, as their dissipation $g_c\,\dot\gamma$ integrates over a load step to the fracture-energy increment $g_c\left[\gamma(\varphi_n) - \gamma(\varphi_{n-1})\right]$. 
The framework thus represents cracks by a smooth field on a fixed mesh, so that the crack path becomes part of the solution. The price is $l_\varphi$ as a second fracture-related parameter and a mesh that must be fine enough to resolve the band.

\subsection{Modeling challenges}
\label{ssec:challenges}
The following challenges arise in practical fracture problems in food production, mastication and beyond. They are introduced here for context and examined in the numerical examples below.
 
\subsubsection{Crack nucleation (sharp interface and phase field)}
\label{sssec:nucleation}
Griffith's criterion governs the propagation of an existing crack. Without a sufficiently strong stress singularity, it cannot predict crack nucleation, though \citep{Francfort1998,kumar_revisiting_2020}. This challenge applies to the sharp-interface and to the phase-field theory. A crack regularized over a finite width $l_\varphi$, however, can nucleate even in intact material. The corresponding nucleation threshold can be interpreted as a critical stress $\sigma_c$ that depends on $l_\varphi$ \citep{Tanne2018,pham_gradient_2011}, so that nucleation is controlled by the numerical regularization rather than by an independent material strength. An explicit relation is available only for simplified cases such as a homogeneously stressed 1D bar,
\begin{equation}
    \sigma_c \propto \sqrt{g_c \, E / l_\varphi} \quad \text{(homogeneous 1D bar)}.
    \label{eq:sigma_c}
\end{equation}
For $l_\varphi \rightarrow 0$, $\sigma_c \rightarrow \infty$ and nucleation is suppressed, as in the sharp-interface limit. To control the nucleation process in the present setups, an initial numerical crack is imposed in both frameworks. Its length is a compromise between a tolerable numerical imperfection and a controlled initiation process, which is physically governed by smaller-scale pre-damage \citep{Tekkaya2020,kumar_revisiting_2020,bikos_effect_2021} or surface roughness \citep{griffith_vi_1921}. For the sharp-interface approach, the initial crack is imposed as a geometric imperfection (disconnected elements). For the phase-field approach, it is imposed by a prescribed damage field (the actual phase field); cf.~\cite{Loiseau_introduce_2025} for the impact of different strategies for imposing the initial crack.

\subsubsection{Free-discontinuity problems (sharp interface)}
In standard sharp-interface simulations, the admissible crack geometries are typically tied to the element edges. The approach is therefore most suitable when the crack path is known a priori from symmetry, specimen geometry or experimental observation such that the mesh topology can be created accordingly. As with a pre-scored chocolate bar that can only break along its grooves, the minimization then reduces to determining the crack extent along this path. 

\subsubsection{Mesh-dependent sharp-interface limit (phase field)}
\label{sssec:meshlimit}
The phase-field width should ideally vanish ($l_\varphi\rightarrow 0$) to approximate the sharp interface. At the same time, it is bounded from below by the element size $h < l_\varphi$ required to resolve the band \citep{linse_convergence_2017,may_numerical_2015,zhang_numerical_2017,nguyen_choice_2016}. Convergence towards the sharp-interface limit can therefore only be tested by refining both numerical parameters together, $l_\varphi$ and $h$ \citep{pandolfi_comparative_2021}.
 
\subsubsection{Curvature interference (phase field)}
\label{sssec:curvature}
Stationarity of Eq.~\eqref{eq:incremental_min_pf} with respect to $\varphi$ involves the term $g_c\,l_\varphi\,\div{\nabla\varphi}$ from the gradient part of $\gamma$. Splitting the phase-field gradient into its magnitude $\norm{\nabla \varphi}$ and direction $\B{n}_\varphi = \nabla \varphi / \norm{\nabla \varphi}$ isolates $\div{\B{n}_\varphi}$, which is the curvature of the phase-field isosurfaces up to a constant factor. For finite $l_\varphi$, the crack curvature thus enters the conditions for crack growth, also see \citep{langenfeld_curvature_2023}. Such curvature effects can be intended, e.g., for deviations from standard fracture in microstructured materials \citep{Saxena2020,Vidyasagar2018,hedjazi_application_2012}, for nucleation \citep{Tanne2018} or for kinking \citep{Li2019}. For a regularization that is merely meant to smear the crack \citep{Peerlings1996,Francfort1998,Dimitrijevic2008, Forest2009, Miehe2010,Kuhn2011}, they are an unintended by-product that only vanishes in the idealized limit case $l_\varphi \rightarrow 0$.
 
\subsubsection{Stress field interference (phase field)}
\label{sssec:stressfield}
Since the stiffness is degraded over a band of width of the order of $l_\varphi$ rather than along a surface, the singular crack-tip stress of linear elastic fracture mechanics is replaced by a finite maximum at a distance of the order of $l_\varphi$ ahead of the band. The sharp crack tip thus becomes a blunted process zone, and the effective strength decreases with increasing $l_\varphi$, cf.~Eq.~\eqref{eq:sigma_c}. Two consequences matter for crack paths. First, the sharp-interface near-tip field is recovered only at distances large compared with $l_\varphi$, so that a hole, notch or free edge within a few $l_\varphi$ interacts with the crack at a distance. Second, crack growth is governed by the elastic energy averaged over a region of size $l_\varphi$, so that stress variations below this scale cannot effectively steer the crack.

%%%%%%%%%%%%%%%%%%%%%%%%%%%%%%%%%%%%%%%%%%%%%%%%%%
%%% Experimental setup %%%
%%%%%%%%%%%%%%%%%%%%%%%%%%%%%%%%%%%%%%%%%%%%%%%%%%
\section{Experimental setup and cooled dark chocolate as a practical case study}
\label{sec:experimentalsetup}

This section provides the experimental basis of the recipe and identifies the elastic parameters before fracture (Step~3).
Cooled dark chocolate is used in this investigation as a food-engineering application for the previously introduced numerical methods. Its mechanical properties depend strongly on composition, crystallization, microstructural heterogeneities, temperature, and loading rate, resulting in a wide range of reported material parameters and associated uncertainties \citep{rezende_mixture_2015,biswas_physical_2017,beckett_science_2019}. For the present analysis, these sources of variability are reduced by considering a single commercially available chocolate type without particulate inclusions and testing it at a low temperature. At the millimeter-to-centimeter scale, the material is assumed to be sufficiently homogeneous and isotropic, allowing the analysis to focus on isotropic fracture behavior. While the pronounced temperature dependence complicates the determination and transferability of elastic, viscous, and plastic parameters, it also enables rapid, resource-efficient, and repeatable specimen fabrication through melting and subsequent cooling. The resulting methodological assumptions and their implications are discussed in the following. 

%%%%%%%%%%%%%%%%%%%%%%%%%%%%%%%%%%%%%%%%%%%%%%%%%%
%%% Available data
\subsection{Available data basis for chocolate}
\label{ssec:choch_data_basis}

Reported mechanical properties vary as widely as the chocolate types themselves.
\citet{Severa2008} determined properties of dark chocolate from compression, bending and splitting tensile (Brazilian) tests, focusing on strain-rate effects and non-linearity, with typical values of Young's modulus $E=9$\,MPa and Poisson's ratio $\nu=0.415$. \citet{bikos_effect_2021,samaras_multiscale_2024} relied on $E \approx 100$\,MPa, while assuming near-incompressibility for the base material of aerated chocolate.
\citet{kumbar_high_2023,kumbar_stress_2024} used split Hopkinson pressure bar tests at room temperature, where plastic and rate effects over frequencies up to 2.5\,MHz are relevant, and obtained $E = 1264.5$--$2303.6$\,MPa and $\nu = 0.4475$--$0.4243$ for extra dark and dark chocolate, respectively. \citet{nedomova_effect_2025} combined both test types and found tensile strengths of 0.5--5\,MPa.
\citet{yang_quantifying_2025} characterized the viscoelastic effect of oleogels as a solid-fat replacement using a fractional Maxwell model ($10^{-2}$--$10^{-1}$\,rad/s, creep times of approximately 100\,s).
\citet{Tremeac2008} showed a strong temperature dependence of Young's modulus in the freezing range ($E$ dropped from $72 \pm 30$\,MPa to $49 \pm 14$\,MPa between $-10$\,°C and +9\,°C), but only a weak one for Poisson's ratio ($\nu$ from 0.39 to 0.40 within $\pm0.06$). \citet{bikos_destructive_2023} confirmed the sensitivity to temperature, rate and loading conditions, with $E \approx 80$--$800$\,MPa, and additionally studied micro-aeration. \citet{zhao_structure-fracture_2018} reported lower fracture stress (2.5--3\,MPa) and Young's modulus (300--600\,MPa) for higher fat content due to reduced particle--particle and particle--fat interaction. For cohesive-zone modeling, \citet{gomez_fracture_2022} used $E = 800$--$1660$\,MPa, maximum stresses of 0.19--0.64\,MPa and fracture energies of 2.1--3.3\,J/m$^2$, depending on sample and notch geometry. A higher limit for fracture energies of 8--22\,J/m$^2$ is reported for milk chocolate simulations in \cite{samaras_combining_2024}.

These values provide a first orientation but do not cover cooled dark chocolate, as compositions, temperatures and/or strain rates differ. The large spread of elastic moduli is attributed mainly to rate dependence, considering its lower end relevant for quasi-static conditions. Crystallization, governed by the six polymorphic forms of cocoa butter obtained by tempering \citep{loisel_phase_1998,gutierrez_stateart_2017}, is another key design and manufacturing parameter and strongly affects rheology \citep{bolliger_comparison_1998,le_reverend_modelling_2009}. Moreover, room-temperature tests tend to show less abrupt crack growth and more plastic deformation than near-freezing temperatures. The parameters were thus derived from extra experiments and checked against the literature, following established quasi-static setups \citep{gomez_fracture_2022,swackhamer_fracture_2019,nedomova_effect_2025}.

%%%%%%%%%%%%%%%%%%%%%%%%%%%%%%%%%%%%%%%%%%%%%%%%%%
\subsection{Sample production}
The chocolate type was chosen among three sorts of dark chocolate (60\,\% chocolate coating sold as Rewe Beste Wahl, 70\,\% chocolate sold as dm bio Feine Bitter Schokolade and 100\,\% chocolate sold as Lindt Excellence Cacao Pur). The second sort was chosen due to local availability and this choice shall by no means be understood as a quantitative or qualitative review of the product or the manufacturer. The chocolate consists of cocoa mass, cane sugar and cocoa butter. The different ingredients and complex crystallization processes can cause chocolate to be a material with a wide range of material values, which is why the following fabrication process was standardized for reproducible material behavior.

Three types of samples were fabricated from the 70\% dark chocolate: bars with thicker ends for clamping in tension experiments, simple bars for compression and three-point-bending specimens (all bars with $5 \times 5\,[\text{mm}^2]$ cross-section), and square blocks ($100 \times 100 \times 5 \,[\text{mm}^3]$) for plate-bending specimens, see Fig.~\ref{fig:experiments} for further details. Acrylonitrile butadiene styrene (ABS) molds were 3D-printed for casting. The molds were first greased and put on an acrylic base plate with baking paper in between. If not otherwise noted, all following production steps were performed at room temperature around 21\,°C.

The chocolate was melted in a 376\,g glass bowl for two minutes at 800\,W inside a microwave. 150--200\,g chocolate (density of approximately 1140\,kg/m$^3$) were broken into 10--12 chunks for one fabrication run. This provides approximately three plate-bending specimens or one plate-bending specimen and three three-point-bending specimens and six tension/compression specimens, including discard. Slow subsequent stirring should check for and avoid air pockets. After pouring the melted chocolate into the molds, the molds were lifted slightly to release possibly undetected air pockets from the lower corners and some chocolate was added again on top to guarantee complete filling. Air pockets were observed in only a few specimens before or after testing (approximately one out of ten) and the respective specimens were discarded.  

After pouring, another layer of baking paper was put on top and another layer of acrylic glass on top of the paper. The paper was slightly stroked from the center to the boundaries and slightly stretched to obtain a planar geometry. 2\,kg weights were finally placed for additional weight. This setup was then allowed to cure for 8--12 hours at approximately 8--10\,°C. Afterwards, the specimens were taken out from the molds with slight and uniform pressure. Releasing them at 8--10\,°C turned out to be beneficial for avoiding plastic deformation or cracking. Still, discard was about 30\,\% at this step and mostly due to imperfect greasing for small specimens.

A hole was subsequently drilled into the plate-bending specimens with a core drill and high feed rate to suppress local heating and measured with a caliper. These samples were then fixed between two acrylic glass plates and slit with a utility knife with a blade with 0.5\,mm thickness. The tension specimens were placed into ABS holders and compression specimens as well as three-point-bending specimens were cut into lengths of approximately 18\,mm and 47\,mm, respectively. All specimens were finally stored in a refrigerator at 3\,°C for at least 6 hours. They were then tested at air temperature of about 15\,°C within one to two minutes after taking them out of the refrigerator.

\subsection{Experiments}
\label{ssec:tests}
 
Four types of tests are performed, each serving one step of the recipe:
\emph{tension} tests of three specimens for Young's modulus (Step~3), \emph{compression} tests of four specimens for a first estimate of Poisson's ratio (Step~3), \emph{three-point-bending} tests of four specimens for the elastic slope and the peak force with a known, straight crack path (Steps~3--5), \emph{plate-bending} tests of four specimens for a freely evolving crack path (Step~6).

Tension, compression and three-point-bending tests were performed on a Kammrath \& Weiss uniaxial testing device with simultaneous picture recording, Fig.~\ref{fig:experiments}. Due to the low force values, its signals were leveled after completion for higher accuracy. The force caused by screwing of clamps was zeroed based on the force level after cracking. Also systematic yet small machine errors, e.g., forces due to erratic thread motion and contact, were subtracted by measurements of a sample-free run performed directly after each test. In the plate-bending tests, a calibrated weight attached by a string to one half was lowered slowly until cracking. Details follow in the sections below for each experiment.

\subsection{Reliability and uncertainties}
The geometric accuracy is limited by imperfections of the mold, the baking paper and the release from the greased mold walls. It was checked on the four three-point-bending specimens, whose $5 \times 5$\,mm$^2$ cross-section is representative of the smallest features of all specimens. Measurements at four points per specimen yield a width of $4.8387 \pm 0.0419$\,mm (confined by the greased mold) and a height of $5.0344 \pm 0.0432$\,mm (confined by baking paper and acrylic plate). These deviations are mainly attributed to the 3D printing and do not grow with specimen size. The plate notches were approximately 1\,mm wide, and the drilled holes had a diameter of 27\,mm with deviations of at most 0.2\,mm. Overall, these deviations are negligible for the present application.

The chocolate is modeled as an ideally brittle, isotropic material. Its linear elastic stiffness degrades with increasing damage. This is justified by the small strains before cracking, the sudden force drop and the clean fracture surfaces without apparent plastic deformation. The three-point-bending tests yield a reproducible peak force of $0.8293~(\pm 0.0586)$\,N. The smooth transition between force increase and drop is attributed to a spatially distributed damage onset in the region of highest stress and to imperfect boundary conditions. Viscous effects are neglected, since the displacement rate of 0.020\,mm/s (strain rate of about 0.001/s) lies at the slow end of the literature data and far below the viscous time scale reported in~\citep{Severa2008}.

The short warming during transport (1--2 minutes) and testing (less than 1 minute) is assumed to be negligible and equal for all specimens. All test specimens were made from chocolate heated only once, and aging, e.g., by fat crystallization, is negligible because of the short, air-tight storage. Specimens heated repeatedly or stored for several days were used only to test the setup, to avoid bias from recrystallization or blooming. A standardized chain of producer, distribution and fabrication supports reproducibility. The derived values are therefore suitable for the present methodological assessment, but the stated deviations should be considered when they are used elsewhere.

%%%%%%%%%%%%%%%%%%%%%%%%%%%%%%%%%%%%%%%%%%%%%%%%%%
\subsection{Derived elastic parameters}

Young's modulus is obtained from tension tests of three specimens, with stresses and strains computed from force and displacement via cross-sectional area and length. Numerical simulations may further support the determination of elastic moduli but are only used later to validate them against the elastic slope from three-point bending. The specimens were clamped in ABS holders, since chocolate otherwise slips already at small forces. With a Young's modulus of about 2\,GPa, i.e., at least a hundred times that of chocolate, the holders' compliance is negligible. Nevertheless, DIC images, uneven crack paths and the force-displacement curves indicate slip at larger forces. This is attributed to partial loss of contact of the chocolate between the clamps and to slight bending of the holders under misalignment, which was clearly observed in one discarded test.

\begin{figure}[h]
  \centering
    \includegraphics[width=0.9\textwidth]{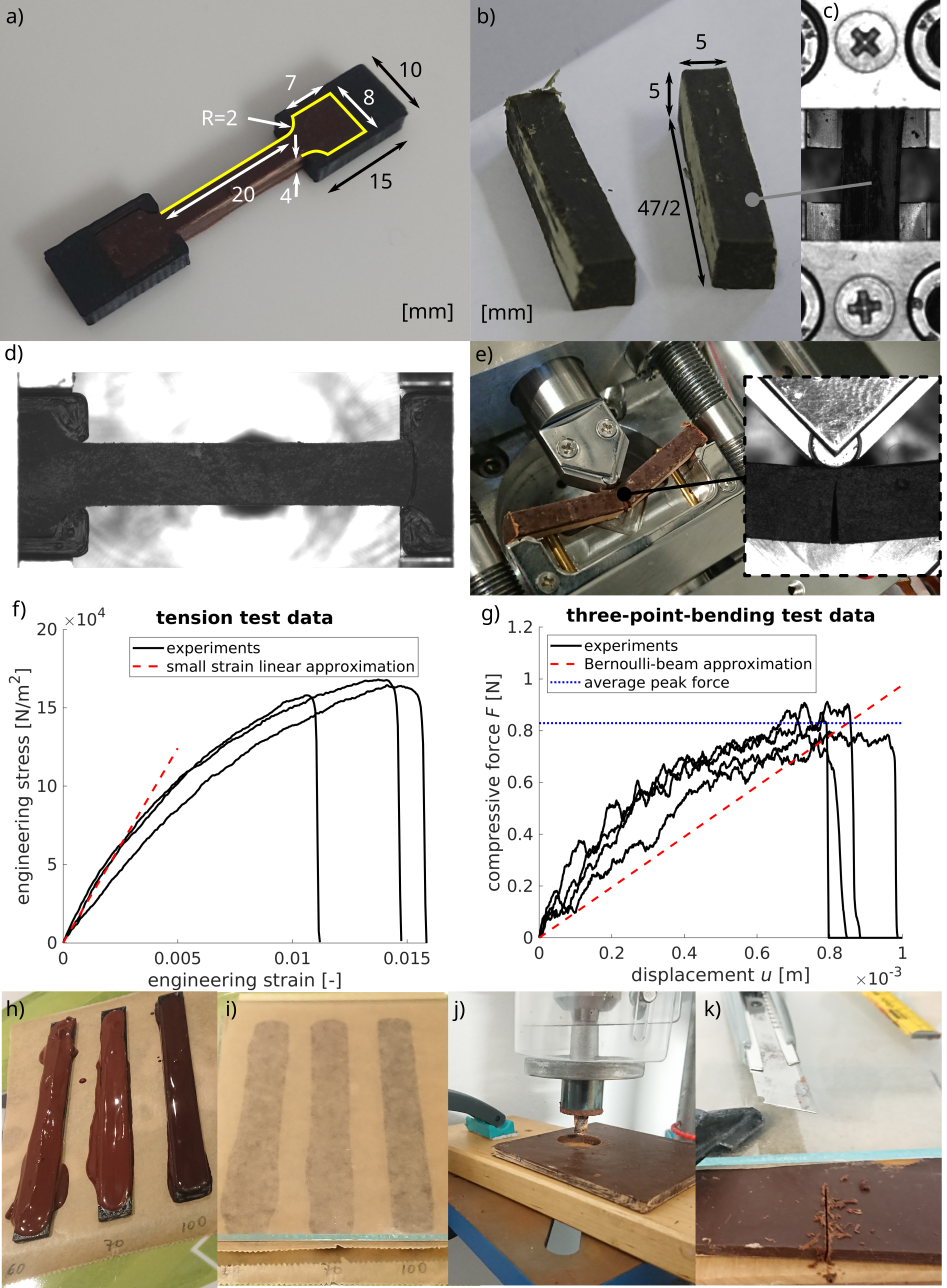}
    \caption{Geometry and results of the parameter identification set-up: a) tension specimen; b) three-point-bending specimen with crack surface (bright coating is residual greasing); c) compression test; d) tension test with crack near the right holder; e) three-point-bending test; f) results of the three tension tests; g) results of the four three-point-bending tests; h) casting of specimen bars for the three sorts of chocolate with 60\,\%, 70\,\% and 100\,\% cocoa content, respectively; i) top cover of the specimen bars; j) drilling of a plate-bending specimen; k) cutting of a plate-bending specimen.
}
    \label{fig:experiments}
\end{figure}

Only the initial, linear part of the tension tests is therefore used (Fig.~\ref{fig:experiments}\,f). A linear regression excludes both the first data points with insufficient resolution and the clamping-induced nonlinearity at larger forces by selecting the ranges with the highest correlation coefficients ($R$ = 0.9965, 0.9975 and 0.9982 for the three samples). Averaging yields $E = 24.844 \pm 6.039$\,MPa, with most of the scatter attributed to the clamping. This value agrees with the lower, quasi-static range of the literature data. Independently, a classical Bernoulli beam approximation \citep{timoshenko_strength_1955} of the linear regime of the three-point-bending tests matches the measurements, confirming the value for a different loading condition (Fig.~\ref{fig:experiments}\,g).

Poisson's ratio is estimated from compression tests of four specimens, made from remnants of the three-point-bending bars and shortened to 18\,mm to prevent buckling. These tests allow only a first estimate, since the cut end faces were not sufficiently plane for even contact and some specimens rotated or slid in the setup. As local DIC strains were not sufficiently sensitive in the lateral direction, absolute pixel length measurements between compressive strains of nearly 0\,\% and 8\,\% were evaluated instead, which may be affected by nonlinearity, warm-up and plasticity. The resulting $\nu = 0.4017 \pm 0.0163$ nevertheless agrees well with the reference value of 0.4 reported in~\citep{Tremeac2008} for a wider temperature range, which is therefore adopted.

In conclusion, the rounded reference values used in the simulations are $E = 25$\,MPa and $\nu = 0.4$.

\clearpage
\clearpage

%%%%%%%%%%%%%%%%%%%%%%%%%%%%%%%%%%%%%%%%%%%%%%%%%%
%%%
\clearpage
\section{Three-point-bending -- Comparison of numerical frameworks for straight cracks}
\label{sec:3pbending}

\begin{figure}[h]
  \centering
  \includegraphics[width=0.95\textwidth]{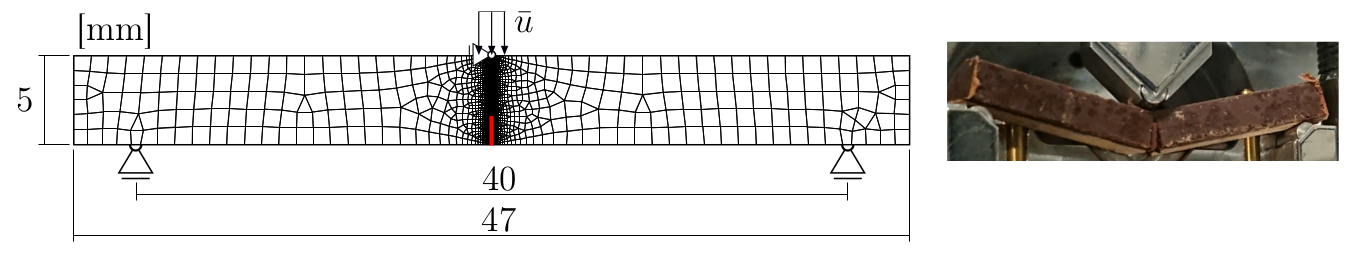}
  \caption{Three-point-bending -- numerical and experimental setup. The length of the imposed loading zone (0.5\,mm, top center) and that of the pre-crack (0.25\,mm, red line at the bottom) are shown not to scale.
  \label{fig:experimental_setup_3pbending}}
\end{figure}

With the elastic parameters fixed, the three-point-bending test (Fig.~\ref{fig:experimental_setup_3pbending} and further numerical settings in \ref{app:numerical_implementation}) serves to identify the fracture-related parameters (Steps~4 and~5). Its practically straight and predictable crack path makes it suitable for both frameworks, the sharp-interface and the phase-field approach. They share the Griffith-type criterion but approximate it differently and therefore differ in their pre- and post-peak behavior. The targets are the elastic slope, $F/u = 0.9766$\,N/mm, and the peak force, $F_{\text{peak}} = 0.83$\,N.

\subsection{Sharp-interface solutions}
\label{ssec:3pbending_sharp}
The simulated elastic slope lies just between the Bernoulli beam theory \citep{timoshenko_strength_1955} and the above-baseline fluctuations of the experiments, which are attributed to the gear tooth contacts and supporting rollers at these low forces. The deviation of 6.6\,\% between the simulations and the theoretical Bernoulli beam approximation is mainly due to the finite slenderness of 40/5 = 8, which prohibits full applicability of the theory, and the reduced stiffness due to the initial crack. The elastic parameters are deliberately not re-fitted to compensate for these side-effects and each parameter remains linked to the loading stage that governs it.

The only fracture-related parameter, $g_c$, is calibrated such that the solution matches $F_{\text{peak}} = 0.83$\,N. A single simulation suffices for calibration, since $F_{\text{peak}} \propto \sqrt{g_c}$ for fixed geometry, crack path and Young's modulus -- rescaling directly yields $g_c = 0.0043$\,N/mm. It acts as a reference for the phase-field simulations below. This value lies between the ranges reported in \cite{gomez_fracture_2022} (0.0021--0.0033\,N/mm) and \citep{samaras_combining_2024} (0.008--0.022\,N/mm). Despite differences in sample and loading conditions, the agreement in order of magnitude supports the suitability of the procedure for characterizing fracture in brittle foods.

Two alternative solution strategies are considered for comparison of the employed local energy minimization, for which a crack propagates if the driving force equals the critical energy release rate. As a first alternative, global energy minimization selects the admissible crack state of lowest total energy, even if it is separated from the current state by an energy barrier \citep{knees_inviscid_2008,mielke_rate-independent_2015,negri_comparative_2010}. It predicts fracture too early, however, at less than half of the experimental force for the same $g_c$. Arc-length control, as the second alternative, can follow the post-peak path by a controlled force-displacement parametrization even for unstable branches. It resolves an unstable snap-back that the other solvers jump across, with the same peak force as the local solution. This snap-back, however, lies beyond the present quasi-static experiment, which is displacement driven only until the specimen breaks by practically instantaneous, uncontrolled crack growth. Afterwards, energy is transformed into kinetic energy and dissipated through free vibrations and the supports. For the present three-point bending of a chocolate beam, local energy minimization is thus not only computationally cheapest but also matches the target values of the elastic slope, the peak force and the prescribed crack path.

\begin{figure}[h]
  \centering
    \includegraphics[width=0.9\textwidth]{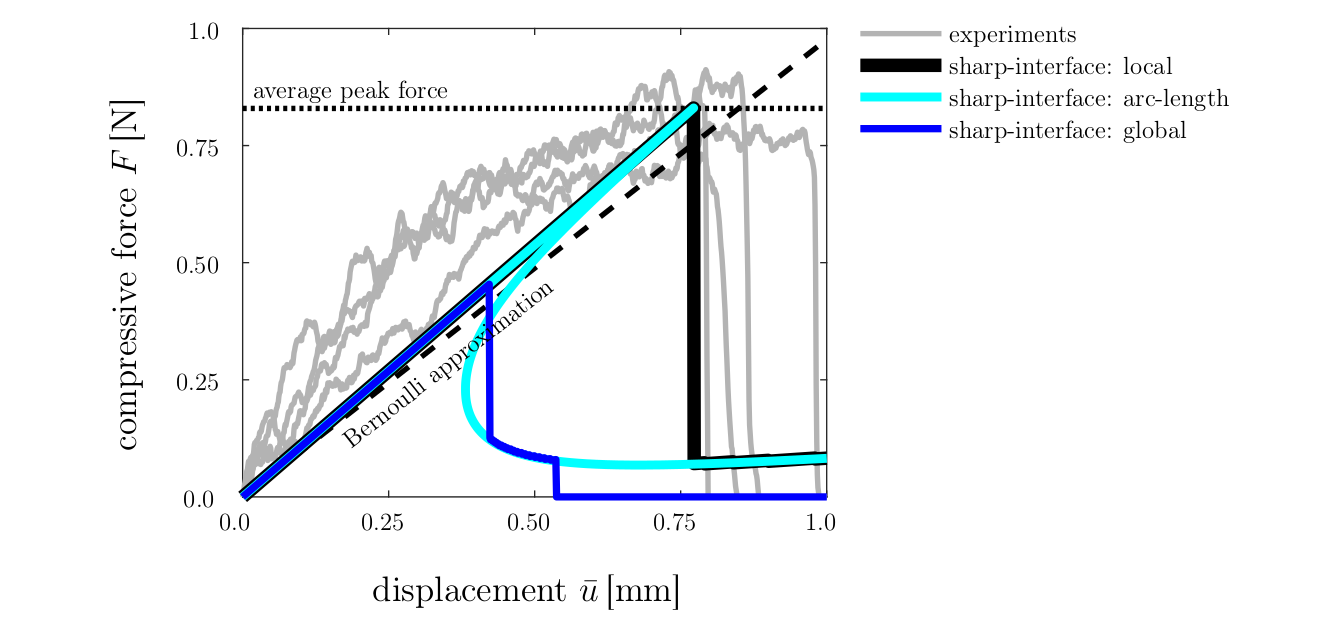}
    \caption{Three-point-bending -- sharp-interface solutions: local, global and arc-length methods of the sharp-interface approach compared to the average peak force and the Bernoulli beam approximation based on the experimentally derived elastic parameters.}
    \label{fig:threepointbending_SharpInterface_Simulations}
\end{figure}

\subsection{Phase-field solutions}
\label{ssec:3pbending_phasefield}

\subsubsection{Numerical calibration and convergence for $l_\varphi$ and $h$}
\begin{figure}[htbp]
  \centering
    \includegraphics[width=1.0\textwidth]{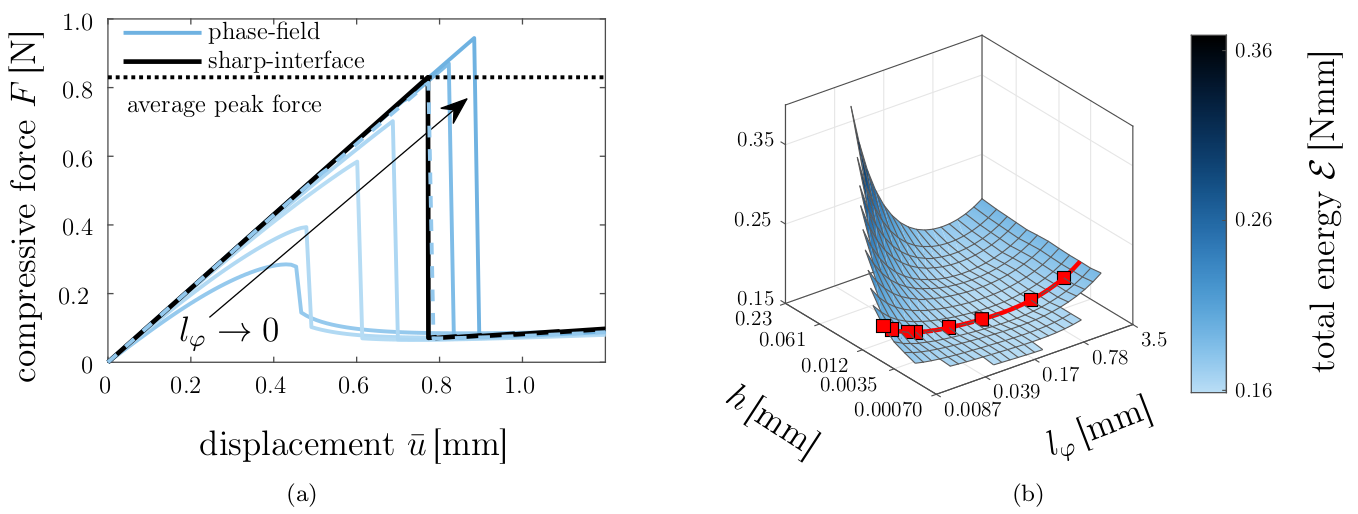}
    \caption{Three-point bending -- phase-field solutions: (a) Structural response for varying phase-field width $l_\varphi$ at fixed mesh size $h$. The dashed line with $l_\varphi = 0.0184$\,mm shows the best match with the sharp-interface approach. (b) Energy landscape for varying $h$ and $l_\varphi$ at fixed $\bar u=1.2$\,mm. Combinations of $l_\varphi-h$ that are computationally infeasible or without a clear pronounced force peak are omitted. The red markers indicate the combinations used for the left plot and the blue color coding of the energy range is transferred to them accordingly. }
    \label{fig:threepointbending_PF_h_vs_lengthscale}
\end{figure}

For comparison, the phase-field simulations first adopt $E$, $\nu$, $g_c$ and the mesh of the sharp-interface simulations, while the phase-field width $l_\varphi$ enters as an additional parameter. Compared with the experimental peak force, the elastic slope and the calibrated sharp-interface solution, $l_\varphi = 0.0184$\,mm provides a good match (dashed line in Fig.~\ref{fig:threepointbending_PF_h_vs_lengthscale} (a)). Smaller and larger $l_\varphi$, however, deviate due to two different types of errors.

Larger values of $l_\varphi$ yield smoother, less instantaneous cracking with lower peak forces \cite{kirkesaether_iterative_2020}. This is due to the regularization error, i.e., the phase-field width becomes too large to match the sharp-interface description in the present experiment. Smaller values of $l_\varphi$ at fixed $h$ do not converge either but overshoot the peak force, in particular towards the theoretically relevant zero-thickness limit $l_\varphi \rightarrow 0$. This is the discretization error, i.e., the phase-field width is no longer resolved by the mesh, which artificially increases the effective fracture toughness \citep{bourdin_variational_2008}.

A rigorous limit therefore requires both $l_\varphi \rightarrow 0$ and $h \rightarrow 0$ with $h/l_\varphi \rightarrow 0$ \citep{bourdin_numerical_2000}, cf.~Sec.~\ref{sssec:meshlimit}. In practice, the appropriate ratio depends on the physical and numerical settings, and common rules of thumb such as $h/l_\varphi \ll 1$, $h/l_\varphi \le 2$ or similar are only orders of magnitude \citep{Miehe2010,pandolfi_comparative_2021}. As a more specific indicator for suitable combinations, $l_\varphi$ can be chosen to minimize the total energy at a given $h$, because the energy reflects both errors, resulting in an energy-minimizing valley in the $l_\varphi$--$h$ plane (see Fig.~\ref{fig:threepointbending_PF_h_vs_lengthscale}~(b) and \citep{linse_convergence_2017} for a detailed methodological discussion). The energy is moreover the natural measure, since $\Gamma$-convergence ensures the convergence of energies and minimizers (but not necessarily of their derivatives such as forces) \citep{Braides2002}. For the present setup, the combination $h = 0.0056$\,mm and $l_\varphi = 0.0184$\,mm is computationally affordable, lies within the energetically reasonable valley and deviates by only $1.9$\,\% from the experimental peak force.

\subsubsection{Mixed calibration with $l_\varphi$ and $g_c$ at finite mesh size $h$}
While the above $l_\varphi$--$h$ combination matches the experimental target values well, including $h$ in the final calibration process is not feasible. Calibration would have to be repeated for every modification of the mesh size $h$ and the limit $h \rightarrow 0$ is restricted by computational resources. Finite values of $h$ and thus of $l_\varphi$ are inevitable. As a consequence, the phase-field width $l_\varphi$ cannot attain the sharp-interface limit and co-determines the structural response. It changes its role from a pure numerical parameter to a parameter interfering with the material model behavior in practice, cf.~\citep{zhang_numerical_2017,nguyen_choice_2016} and \citep{linse_convergence_2017,may_numerical_2015} for methodological guidelines and discussions. A practical solution is to treat $g_c$ and $l_\varphi$ both as fracture-related parameters and to calibrate them together.

\begin{figure}[htbp]
  \centering
    \includegraphics[width=1.0\textwidth]{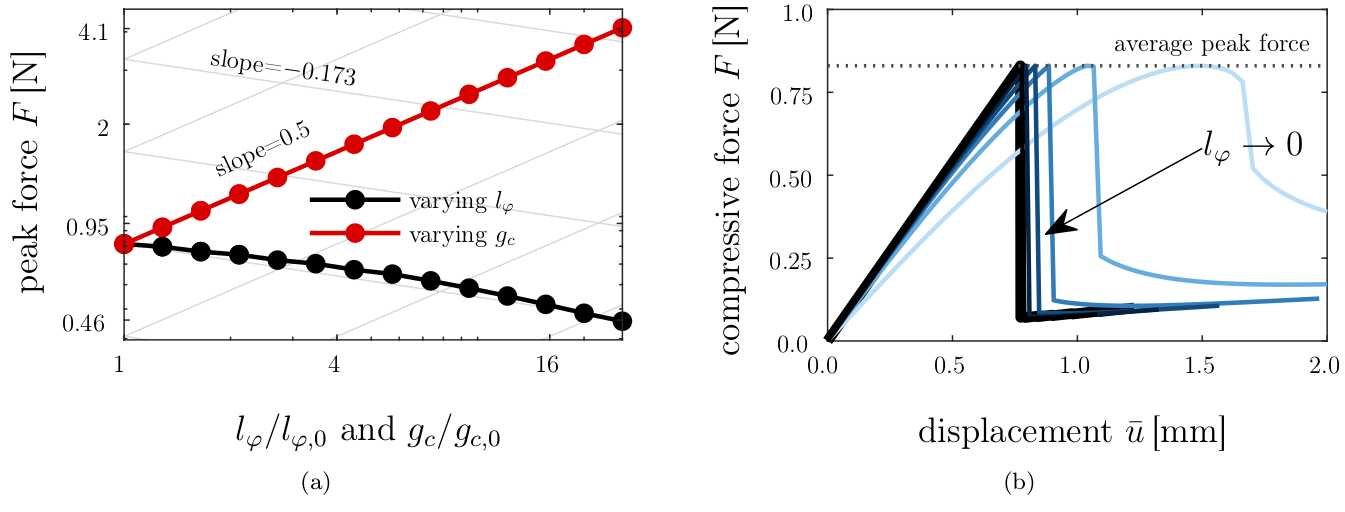}
    \caption{Three-point bending -- phase-field solutions: (a) Double-logarithmic peak force over dimensionless variations of $g_c$ and $l_\varphi$ around the standard settings. (b) Structural responses for decreasing phase-field widths $l_\varphi$ with rescaled $g_c$ such that $g_c^{1/2} \, l_\varphi^{-0.173}=\text{const.}$}
    \label{fig:threepointbending_PF_gc_vs_lengthscale}
\end{figure}

Both parameters have a significant yet opposed influence on the peak force (Fig.~\ref{fig:threepointbending_PF_gc_vs_lengthscale}a). On the one hand, the variation of $g_c$ reproduces the exponent $1/2$ as expected from the sharp-interface investigation. On the other hand, the variation of $l_\varphi$ affects the peak force by the exponent $-0.173$, yielding:
\begin{equation}
    F_{\text{peak}} \propto g_c^{1/2} \, l_\varphi^{-0.173}
    \quad
    \text{(present three-point bending setting)}.
    \label{eq:scaling_3pb}
\end{equation}
This behavior is different from the scaling $g_c^{1/2} \, l_\varphi^{-1/2}$ assumed for a homogeneous stress state in Eq.~\eqref{eq:sigma_c}. \citet{Tanne2018} showed that this different sensitivity to $l_\varphi$ is governed by the stress concentration at the nucleation site: a non-singular stress field recovers the exponent $-1/2$, whereas an ideal crack-tip singularity yields pure Griffith behavior independent of $l_\varphi$. The present specimen contains an initial crack at finite mesh resolution and is consistently an intermediate case.

The final calibration via $g_c$ and $l_\varphi$ then requires a single step. First, $l_\varphi$ is chosen within its admissible range, e.g., near the energy minimum for the given $h$. Then $g_c$ is rescaled to maintain the experimental peak force; here via $g_c \propto (l_\varphi^{+0.173})^2$ according to Eq.~\eqref{eq:scaling_3pb}. This matches the experimental peak force by construction (Fig.~\ref{fig:threepointbending_PF_gc_vs_lengthscale}b) and avoids an adaptation of the mesh size $h$. 

Matching the peak force alone, however, does not yet reproduce the entire structural response. In addition to the $l_\varphi$--$g_c$ calibration, $l_\varphi$ must remain small enough to meet the instantaneous fracture profile (Fig.~\ref{fig:threepointbending_PF_gc_vs_lengthscale}b). Otherwise, the force peak becomes overly smooth, resulting in a biased elastic slope and peak displacement. Especially the distinct peak displacement matters for displacement-controlled processes such as grinding, kneading and, in part, chewing. Finally, $l_\varphi$ must be sufficiently small, the mesh must resolve it, and the fitted $g_c$ is a parameter of the chosen regularization.

%%%%%%%%%%%%%%%%%%%%%%%%%%%%%%%%%%%%%%%%%%%%%%%%%%
%%% 
\section{Plate-bending -- phase-field width and crack path interference}
\label{sec:platebending}

In the previous three-point-bending experiment, the crack path was practically pre-determined: the stress singularity governed crack nucleation, while the symmetry of the problem constrained its subsequent propagation. In many food products, crack paths are likewise intentionally or unintentionally influenced by the specimen geometry or by inclusions. This experiment shifts the focus from a pre-defined straight crack to an unknown, freely evolving crack path. This situation is of equal or even higher importance, for example, for the transportation of chocolate, chips or eggs.  

The following experiment considers a free crack through a notched plate of cooled dark chocolate subjected to an out-of-plane bending load, cf.~Fig.~\ref{Fig:experimental_setup}. The initial notch and a hole drilled on one side of the plate define controlled start and end points for the crack. The path between these points, however, is not prescribed by grooves or other geometric features. Instead, it is allowed to evolve freely and is dictated solely by the interplay of elasticity and brittle crack growth.

The phase-field method is employed for the numerical simulations, since its mesh-independent formulation makes it particularly suitable and robust for freely evolving crack paths \citep{Langenfeld2022}. Although it is a consistent approximation of Griffith-type fracture, its finite width interferes with crack curvature and stress fields \citep{langenfeld_curvature_2023,Li2019}, as motivated in Secs.~\ref{sssec:curvature} and~\ref{sssec:stressfield}. Accordingly, this example deliberately focuses on how the freely evolving crack path observed in the experiment compares with phase-field simulations employing different phase-field widths $l_\varphi$, while keeping the other parameters fixed (Step~6). Recalibration of the other parameters would change the elastic slope or peak limits and mask or bias the original interference in this assessment.

%%%%%%%%%%%%%%%%%%%%%%%%%%%%%%%%%%%%%%%%%%%%%%%%%%
%%% Setup
\subsection{Setup}
\begin{figure}[h]
  \centering
  \includegraphics[width=\textwidth]{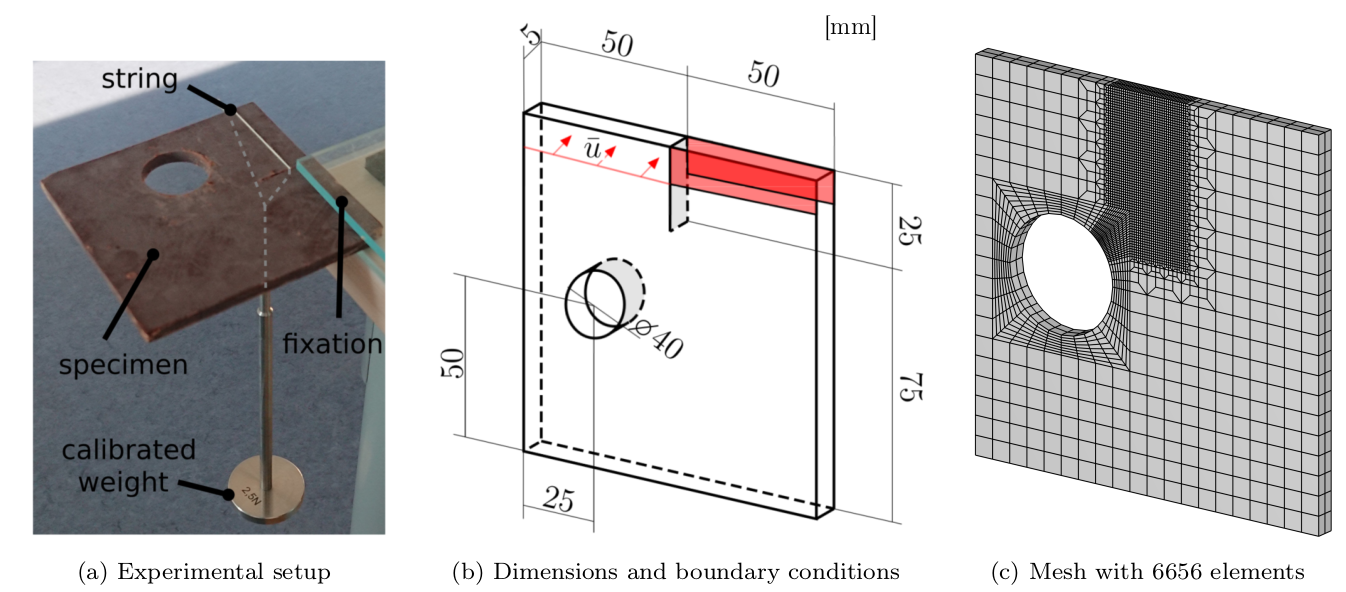}
  \caption{Plate-bending: experimental setup, geometry and numerical mesh.
  \label{Fig:experimental_setup}}
\end{figure}

For the present analysis of chocolate, a plate of 100\,mm $\times$ 100\,mm $\times$ 5\,mm was cast, drilled and notched as shown in Fig.~\ref{Fig:experimental_setup}. One half separated by the notch was clamped. A calibrated weight was attached by a string to the other half and lowered slowly to induce cracking that occurred for forces between 5.0\,N and 7.5\,N. The material used to capture brittle behavior is the same cooled dark chocolate with 70\,\% cocoa at 3\,°C. 
The numerical mesh is refined around the pre-crack (the notch) and on the right side of the drilled hole to sufficiently resolve the expected cracking path. The calculations are performed in a three-dimensional setting, see~\ref{app:numerical_implementation} for further details.

\subsection{Results and discussion}

\begin{figure}[h]
  \centering
    \includegraphics[width=\textwidth]{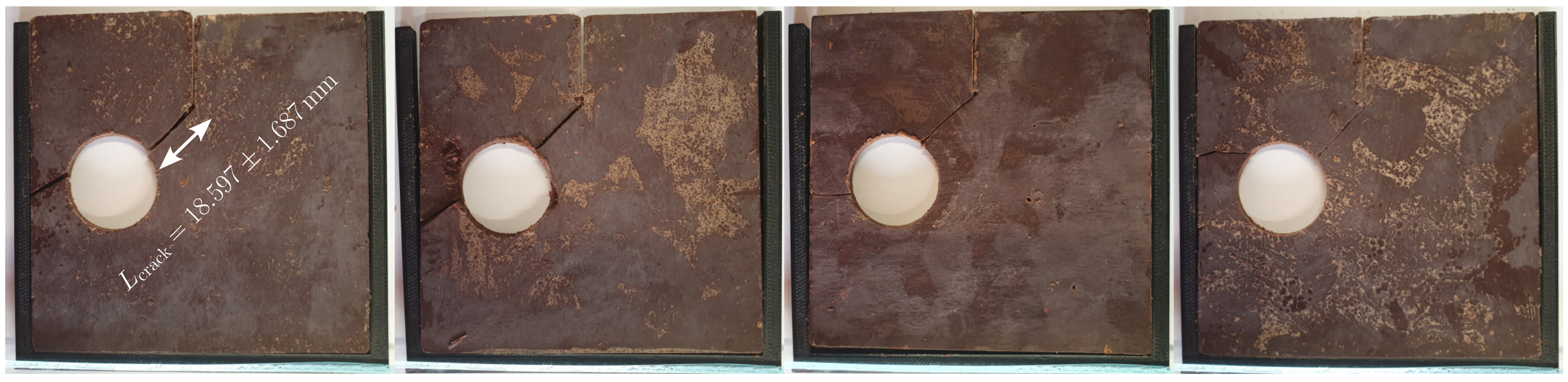}
    \includegraphics[width=\textwidth]{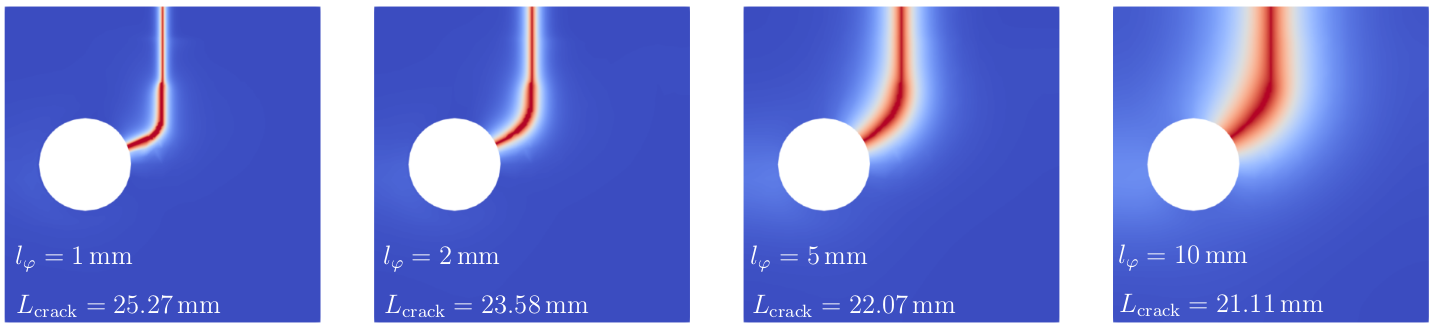}
  \caption{Plate-bending: Experimental specimens after cracking (top) and numerical results for varying phase-field widths $l_\varphi$ (bottom). \label{Fig:CrackingExperiment_ExperimentalNumericalResults}}
\end{figure}

All four specimens show a very similar, straight crack path, Fig.~\ref{Fig:CrackingExperiment_ExperimentalNumericalResults}. The crack length is $18.597 \pm 1.687 $\,mm (mean $\pm$ standard deviation). The crack shape resembles a straight line with a coefficient of determination $r^2_{\text{det}} = 0.983$. This underlines that the notch and the hole reliably govern the crack evolution and yield reproducible results despite the fast propagation process and inaccuracies in fabrication and boundary conditions. Note that the relatively small experimental deviations result from variations in the crack paths and also from fabrication inaccuracies affecting the notch-hole distance.

\begin{figure}[h]
  \centering
    \includegraphics[width=1.0\textwidth]{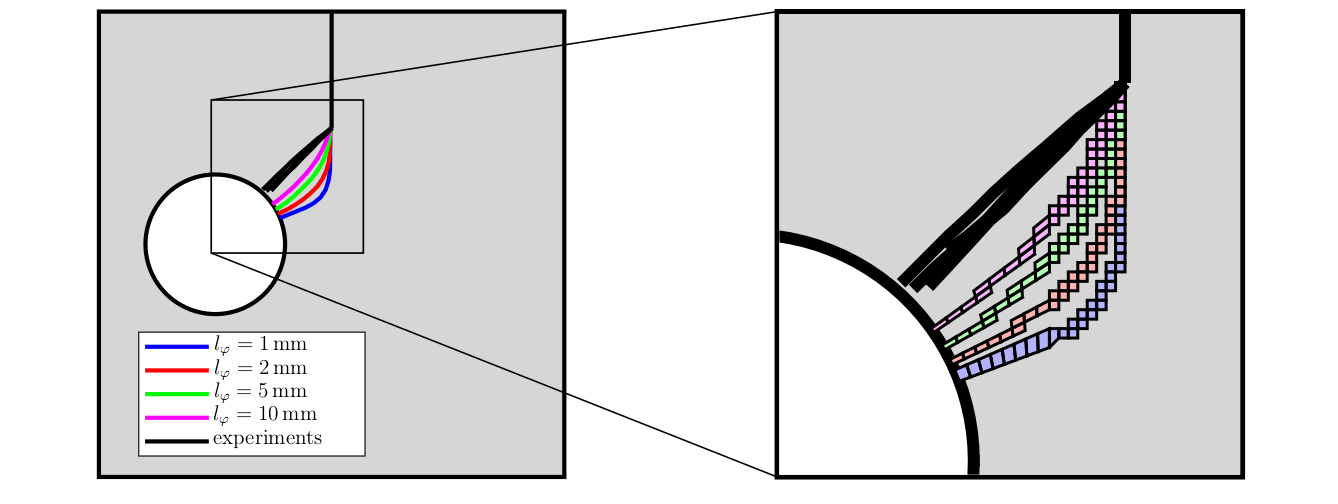}
  \caption{Plate-bending: comparison of the crack path between simulation and experiments. \label{Fig:CrackingExperiment_ExpVsSim}}
\end{figure}

The simulations approach the experimental observation with larger phase-field widths, see Fig.~\ref{Fig:CrackingExperiment_ExperimentalNumericalResults}. This agrees well with the fact that a larger phase-field width penalizes curvature and supports the interaction between larger stress fields and notches. Agreement between experiment and simulations -- in terms of crack length -- is closest for the largest phase-field width involved ($l_{\varphi} = 10$\,mm) with a relative deviation of $13.9$\,\%. The deviation increases up to $33.1$\,\% for the smallest simulated phase-field width ($l_{\varphi} = 1$\,mm). This match between experiment and simulations also applies to the overall crack shape, Fig.~\ref{Fig:CrackingExperiment_ExpVsSim}. Quasi-static modeling, numerical approximations and experimental inaccuracies preclude a perfect match, and the curvature and stress-field contributions cannot be isolated individually. Nevertheless, $l_\varphi$ clearly influences the predicted path in this setup. 

Interestingly, the small $l_\varphi$ limit evolves towards the hole at later stages of crack progression and thus tends to poorer predictions for this problem.
This preference for large $l_\varphi$ is opposite to the three-point-bending result, where a small $l_\varphi$ was required to approach the sharp-interface response and to limit the shift of the peak displacement (Sec.~\ref{ssec:3pbending_phasefield}). Both examples thus reflect a trade-off that cannot be resolved by the phase-field width alone: smaller values better reproduce the singular near-tip field and approach the Griffith limit, but increase the sensitivity to local stress concentrations and reduce the curvature penalization. Larger values smooth the stress field and stabilize the path, but shift the peak displacement, widen the damage zone and may suppress relevant geometric features. Peak force, peak displacement and crack path can therefore favor different ends of an admissible range that is bounded from below and above by practical restrictions like mesh size, and $l_\varphi$ must be chosen with respect to the objective of the application.

\clearpage
\section{Modeling recipe for brittle fracture in foods by the practical example
         of cooled dark chocolate}
\label{sec:recipe}

The preceding sections examined the sharp-interface and the phase-field framework for fracture in foods parameter by parameter. This section condenses the findings into a modeling recipe, i.e., an ordered sequence of decisions and identification steps that can be followed with the computational resources typically available in an applied setting, see Fig.~\ref{Fig:Recipe} for the initially motivated order. Its guiding principle is that numerical and material model parameters are not independent at finite mesh size and must therefore not be identified independently.

\subsection{The six steps}
\label{ssec:recipe_steps}

\emph{Step 1: framework}. The decisive criterion is the information on the crack path. If the path is known in advance, e.g., from scoring as in a pre-grooved chocolate bar, from symmetry, from a perforation line in a packaging film or from observation, the sharp-interface description is the more direct and economical choice, with $g_c$ as the only fracture-related parameter. If the path is part of the solution, the phase-field description is preferable, at the price of $l_\varphi$ and a mesh resolving its width.

\emph{Step 2: discretization and initial crack}. The element size $h$ and the initial crack can practically affect the structural response. The element size balances accuracy and computational budget and, for the phase field, later limits the admissible range of $l_\varphi$. An initial crack makes propagation and hence $g_c$ the governing mechanism.

\emph{Step 3: elastic parameters}. $E$ and $\nu$ are identified before fracture, where no fracture-related parameter is active, and cross-checked against an independent configuration. Where clamping is unreliable, alternative boundary conditions are preferable.

\emph{Step 4: critical energy release rate}. $g_c$ is identified from a specimen with a known crack path. Since $F_{\text{peak}} \propto \sqrt{g_c}$ for fixed geometry, crack path and stiffness, a single simulation and rescaling suffice. This scaling is the practical core of the recipe and also holds for the phase field.

\emph{Step 5: phase-field width and rescaled $g_c$}. For the phase-field approach, $l_\varphi$ is chosen within its admissible range and $g_c$ is rescaled to the peak force in one step. The exponent of $l_\varphi$ must be determined for the problem at hand; food products with notches, pores, corners or scoring lines are closer to the pre-cracked case than to the homogeneous one. The resulting $g_c$ belongs to the regularization and must be reported together with $l_\varphi$ and $h$. Eventually, the force-displacement profile including the peak displacement must be checked.

\emph{Step 6: verification against final crack path}. The parameter set is confronted with the final boundary value problem, potentially with a freely evolving crack. Its path must be checked for the sharp-interface approach (for compatibility with mesh topology) and for the phase-field approach (for possible curvature and stress field interference).

\subsection{Implications for food engineering}
\label{ssec:recipe_food}

Several aspects of the procedure are specific to food applications rather than
generic fracture mechanics.

\emph{Force levels and measurement.} Peak forces of the order of a few newtons are common for brittle foods and lie far below the range for which universal testing machines are usually operated. Appropriate load cells and signal leveling by sample-free reference runs can be advisable.

\emph{Force- versus displacement-controlled processes.} Handling, packaging and transport are force-controlled, and the peak force is decisive. Grinding, kneading and, in part, chewing are closer to displacement control, where the peak displacement matters. The objective must be chosen before the calibration.

\emph{Sources of material variability.} Composition, crystallization state, aging and segregation affect the mechanical response. Chocolate allowed specimens to be produced with various geometric features, but a standardized fabrication route is a prerequisite, which kept the geometric scatter around $1$\,\% and the peak force reproducible within $7$\,\%.

\emph{Crack path as a product property.} In mechanical and civil engineering, the design objective is usually to avoid fracture and eventual failure altogether. An aircraft component or a load-bearing structure is dimensioned such that the critical state is not reached within the service life, and safety margins against crack initiation are of higher concern than the details of the post-critical evolution. In many food applications the situation is reversed, since fracture is intended and routinely provoked during cutting, grinding or mastication. The crack path is then not a numerical detail but a product property in its own right, determining fragment geometry, processability, aroma release and perceived quality. Scoring lines, inclusions, pores and layered structures make such paths partly predictable, which is precisely where the sharp-interface framework is the efficient choice and why both frameworks have been compared in this study.

\section{Conclusion and outlook}\label{sec:conclusion}

This work assessed how two well-established numerical frameworks for Griffith-type brittle fracture, a sharp-interface approach and a diffuse phase-field approach, can be calibrated to predict fracture in brittle food with realistic computational resources. The present investigation is intended as a practical assessment of the numerical modeling capabilities relevant to food engineering, in combination with a physically rigorous calibration and interpretation of the numerical and the material model parameters. To this end, it deliberately explores numerical parameter variations that have not yet been considered in this application, using cooled dark chocolate as the model material. It can be cast into reproducible, non-standard geometries with little effort and remelted for reuse, and a standardized fabrication route kept the scatter in specimen dimensions small.

The findings are structured along a modeling recipe (Fig.~\ref{Fig:Recipe}) and eventually condensed for brittle fracture in foods in Sec.~\ref{sec:recipe}. The choice of the numerical framework is not a numerical detail. The sharp-interface approach reproduced the elastic slope, the peak force and the straight crack path of the bending tests with the lowest computational effort. However, it is most suitable for crack paths known a priori. The phase-field approach is more suitable for free-discontinuity problems, but at the cost of the phase-field width $l_\varphi$ being required as an additional parameter. For a finite mesh size of the simulation, the calibrated parameters are however tied to the numerical setting and its regularization rather than being individual material constants. As a result that is not compatible with idealized theoretical assessments, numerical and material model parameters interfere. The critical energy release rate could not be transferred to the phase-field model, for example, because the phase-field width co-determines the structural response. Scaling relations for the model parameters derived for idealized homogeneous settings moreover cannot be applied without scrutiny to food products with notches, pores or sharp corners.
Most consequential in practice, the calibration targets can conflict. Matching the peak force with a finite $l_\varphi$ can still leave the peak displacement shifted, and the crack path of the plate favored larger values while the bending response favored smaller ones. A parameter set is consequently valid for a stated objective and a stated boundary value problem, and both have to be reported alongside it for non-idealized limit cases.

Foods often contain inclusions, pores or layered structures and respond sensitively to temperature and loading rate \cite{dammass_ratedependent_2024}, which motivates future extensions of the presented framework. During mastication, for instance, heating can turn a glassy, brittle food ductile \citep{schuldt_high-speed_2018,vanderbilt_assessment_2011} and biscuits, nuts or frozen products can add further complexity. Coupling the fracture models with temperature \cite{miehe_thermo-elastic_2015}, plasticity \citep{sprave_large_2023}, heterogeneity \citep{boddecker_effect_2023}, anisotropy \citep{feike_accuracy_2025}, controlled curvature dependence \citep{langenfeld_curvature_2023} and dynamic formulations for post-peak fragmentation are hence natural next steps. Experimentally, alternative tests such as shearing or the splitting tensile (Brazilian) test are preferable where clamping becomes unreliable, the latter at the price of a non-uniform stress state \citep{Severa2008,nedomova_effect_2025}.

For the practice of food engineering, the present findings mean that the objective has to be fixed before the calibration, not after it. Which of the three targets governs follows from the process, as discussed in Sec.~\ref{ssec:recipe_food}: food products differ from load-bearing structures in that fracture is often intended rather than avoided, so the crack path can be the design objective itself rather than a failure to be excluded. The recipe in Fig.~\ref{Fig:Recipe} provides an order to guide the choice of the numerical framework and the parameter identification process.

\section*{Acknowledgements}
Financial support from the German Research Foundation (DFG) via SFB/TRR 188 (project number 278868966, project C01) and via SPP 2256 (project number 441222077, project 12) is gratefully acknowledged. Furthermore, the authors gratefully acknowledge the computing time provided on the Linux HPC cluster at TU Dortmund University (LiDO3), partially funded in the course of the Large-Scale Equipment Initiative by the German Research Foundation (DFG) as project 271512359. The authors would also like to thank Henning Lammen and Stavros Bozinis for the support and discussions on the three-point-bending simulations.

\bibliographystyle{elsarticle-harv} 
\bibliography{chocolate.bib}

%%%%%%%%%%%%%%%%%%%%%%%%%%%%%%%%%%%%%%%%%%%%%%%%%%
%%% Appendices
%%%%%%%%%%%%%%%%%%%%%%%%%%%%%%%%%%%%%%%%%%%%%%%%%%
\clearpage

\appendix

\section{Numerical implementation}
\label{app:numerical_implementation}
The simulations are based on a finite-element implementation in Matlab, employing the parameter values as summarized in Tab.~\ref{Tab:Parameter_Chocolate} if not otherwise noted.

\begin{table}[h]
\centering
\begin{tabular}{lcl|l}
  \hline \hline
  Name                      & Symbol      & Unit 	  & Value          \\
  \hline \hline 
  Young's modulus           & $E$         & [MPa]	  & 25           \\
  Poisson's ratio           & $\nu$       & [--]	      & 0.4            \\ 
  residual stiffness factor & $c_0$       & [--]       & $10^{-7}$      \\ \hline
  damage threshold          & $\kappa_0$  & [MPa]      & 0              \\ 
  crit. energy release rate         & $g_c$       & [N/mm]       & 0.0043        \\ \hline
  phase-field width     & $l_\varphi$ & [mm]       & varying        \\ \hline \hline
\end{tabular}
\caption{Material and numerical parameters used for the simulations if not otherwise specified.
\label{Tab:Parameter_Chocolate}}
\end{table}

\subsection{Sharp interface}  
In the finite-element implementation of the sharp-interface approach, $\Gamma$ is a union of element edges (faces in 3D) across which no tractions are transmitted. Once an edge becomes part of $\Gamma_n$, its nodes are duplicated and each copy is assigned to the element on one side, which permits a displacement jump, cf.~\citep{Giacomini_discontinuous_2003,knees_computational_2013,knees_fully_2022}. No further degrees of freedom are introduced, and each element remains purely elastic. The minimization~\eqref{eq:incremental_min} then runs over a finite set of candidate crack sets $\Gamma \supseteq \Gamma_{n-1}$: for each candidate, the elastic equilibrium problem is solved, and the configuration with the lowest total energy is selected. Due to the quadratic scaling of the bulk energy in terms of the prescribed displacements, only one finite-element computation is required for each crack configuration.

For the three-point bending setup, the beam is discretized in two dimensions by 31k bilinear quadrilateral and triangular elements. A structured mesh of square bilinear elements with edge length $h = 0.0056$\,mm resolves the region around the symmetry line. The element edges along this line define the admissible crack set, i.e., the prescribed crack path. Plane stress conditions are applied and the pre-crack of length $0.25$\,mm is prescribed by means of disconnected elements. 

\subsection{Phase field}
The phase-field simulations share the same material parameters and mesh topologies as the sharp-interface approach. Peculiarities and differences are stated in the following. The displacement field and the phase field are both discretized by linear shape functions. Irreversibility is enforced by the history field approach of \citet{Miehe2010}. The incremental problem~\eqref{eq:incremental_min_pf} is solved for a local solution using the algorithm in \citep{roerentrop_etal_2026}. The initial pre-cracks are prescribed by a fully damaged field, setting $\varphi = 1$. Damage at the supports and in the loading region of the three-point bending specimen is moreover prevented by Dirichlet-type boundary conditions for the phase field with $\varphi = 0$. For the plate-bending setup, two layers of linear hexahedral elements in thickness direction ($\sim 7$k in total) capture the bending behavior of the experiment as a compromise between accuracy and calculation time.

\end{document}